%% file: ferron_dynamics.tex
\documentclass[pra,aps,twocolumn,floatfix,
               showpacs,noshowpacs,
               superscriptaddress, % Puts affiliations with footnotes.
]{revtex4-1}
\usepackage{amsmath}
\usepackage{amssymb}
\usepackage{epsfig}
\usepackage{multirow}
\usepackage{bm}
\usepackage{color}
\usepackage{hyperref}
\usepackage{url}

\input{newcommands}
\usepackage{color}

\begin{document}
% \preprint{NT@UW-12-17}

\title{
Dynamics of spin-polarized impurity in ultracold Fermi gas
}

% Authors: Alphabetical order
\author{Piotr Magierski}\email{piotrm@uw.edu}
\affiliation{Faculty of Physics, Warsaw University of Technology, Ulica Koszykowa 75, 00-662 Warsaw, Poland}
\affiliation{Department of Physics, University of Washington, Seattle, Washington 98195--1560, USA}

\author{Bu\u{g}ra T\"uzemen}\email{bugra.tuzemen@pw.edu.pl}
\affiliation{Faculty of Physics, Warsaw University of Technology, Ulica Koszykowa 75, 00-662 Warsaw, Poland}

\author{Gabriel Wlaz\l{}owski}\email{gabriel.wlazlowski@pw.edu.pl}
\affiliation{Faculty of Physics, Warsaw University of Technology, Ulica Koszykowa 75, 00-662 Warsaw, Poland}
\affiliation{Department of Physics, University of Washington, Seattle, Washington 98195--1560, USA}

\begin{abstract} 
We show that the motion of spin-polarized impurity (ferron) in ultracold atomic gas
is characterized by a certain critical velocity which can be traced back
to the amount of spin imbalance inside the impurity. We have calculated the 
effective mass of ferron in two dimensions. We show that the effective mass scales with the surface 
of the ferron.
%and in general it scales as $M_\mathrm{eff} \propto R^{D-1}$, where $D$ is the dimensionality  of the system. 
We discuss the impact of these findings; in particular, we demonstrate 
that ferrons become unstable in the vicinity of a vortex.
\end{abstract}

%67.85.De	Dynamic properties of condensates; excitations, and superfluid flow
%67.85.Lm	Degenerate Fermi gases
%74.40.Gh	Nonequilibrium superconductivity
%74.45.+c	Proximity effects; Andreev reflection; SN and SNS junctions 
%74.50.+r	Tunneling phenomena; Josephson effects (for SQUIDs, see 85.25.Dq; for Josephson devices, see 85.25.Cp; for Josephson junction arrays, see 74.81.Fa)

\maketitle

\section{\label{sec1}Introduction}
The ultracold atomic gases with nonzero spin polarization offer the possibility to investigate 
the existence of metastable structures that may spontaneously occur in such systems. 
These include realizations of the Fulde-Ferrell-Larkin-Ovchinnkov phase (FFLO)~\cite{ff,lo} leading to the possible formation of liquid crystals~\cite{radzihovsky_liquid}, supersolids~\cite{bulgacforbes}, 
which also include polarized vortex cores~\cite{drummond, inotani, magierski_vortex}, and the Sarma phase \cite{sarma,gubbels,wilczek}.
Although the experimental confirmation of these phases is still lacking, the progress in experimental 
techniques allow to treat spin imbalance as a controllable experimental "knob" and, thus,
offers the possibility to investigate the superfluid gas as a function of 
spin polarization~\cite{zwierlein, partridge, shin, nascimbene}. In particular, the evolution 
of spin-imbalanced systems from the deep BCS regime through the unitary limit to the Bose-Einstein condensate side is 
predicted to generate various exotic phases \cite{sheehy, sheehy_magnetized, radzihovsky_liquid1}.
Although the phase diagram as a function of spin-polarization remains still
merely a theoretical prediction, yet another question may be posed: 
Does the ultracold atomic gas with nonzero spin polarization
admit the presence of metastable structures inside the superfluid, where the polarization
could be effectively stored?
One such structure in the form of a {\it ferron}, resembling the Larkin-Ovchinnikov droplet, has
been recently investigated in Refs. \cite{ferron1,ferron2}. In this case, it was found that one can 
generate dynamically the local spin imbalance in the form of a droplet in an otherwise unpolarized medium 
corresponding to the unitary Fermi gas.
Due to the particular nodal structure of the pairing field, the ferron appears as 
an excitation mode of a metastable character.
On the other hand, one may expect that under the condition of nonzero spin imbalance 
spatially separated ferrons may appear spontaneously 
in the cooling process. This situation may occur in the limit when the 
spin imbalance is too small to generate the FFLO phase in the bulk.
%&&&&&&&&&&&&&&&&&&&&&&&&&&&&&&&&&&&&&&&&&&&&&&&&&&
\begin{figure}[b]
   \begin{center}
   \includegraphics[width=1.2\columnwidth, trim=80 120 0 0, clip]{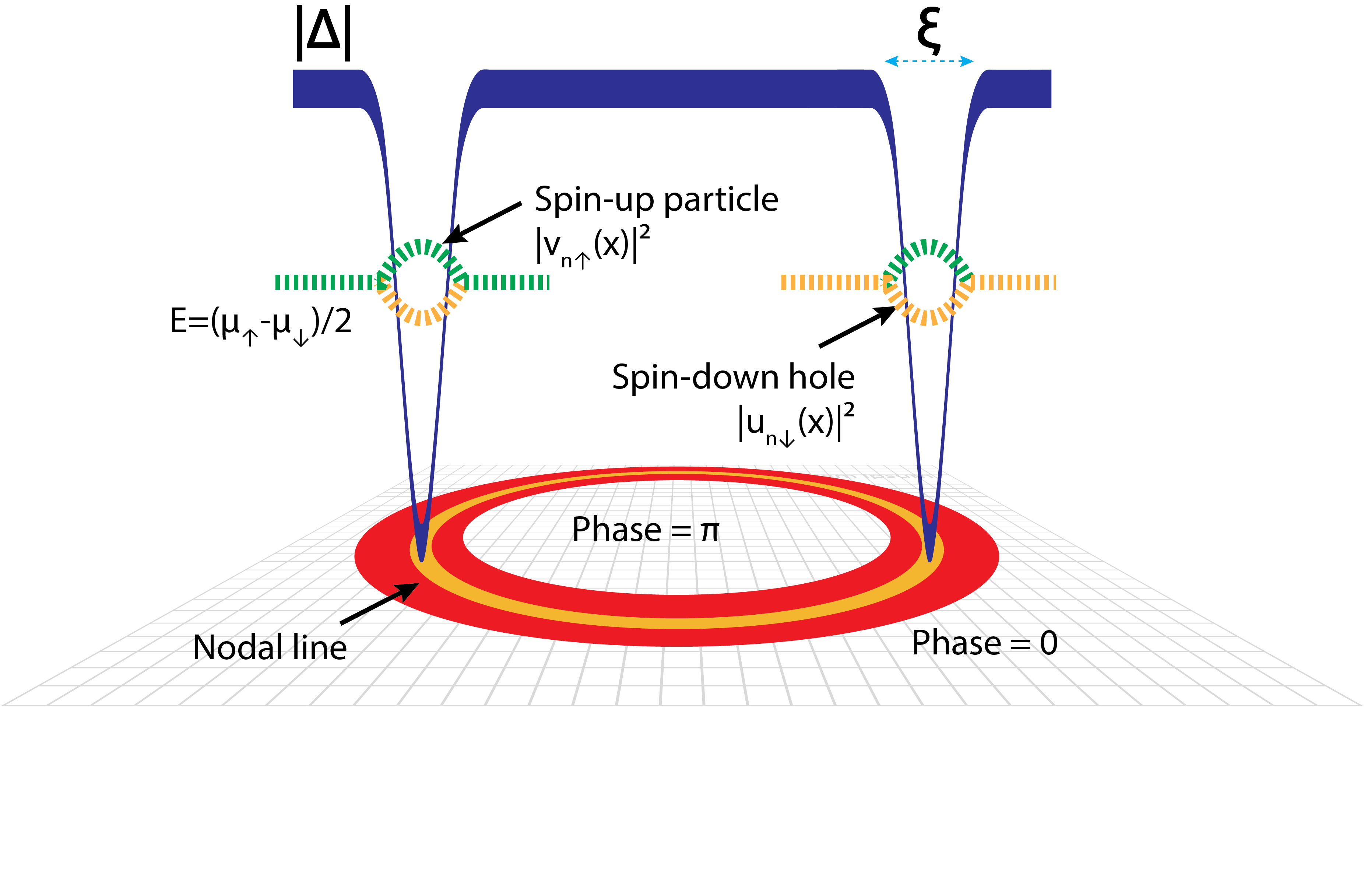} 
   \end{center}\vspace{-3mm}
   \caption{Schematic presentation of the ferron structure in two dimensions (2D). It is characterized by a nodal line where the order parameter $\Delta$ changes phase by $\pi$. The localized Andreev states reside around the nodal line (red area) and they accumulate majority (spin-up) particles. These states are almost degenerate, with excitation energies about $\frac{\mu_\uparrow-\mu_\downarrow}{2}$. }
   \label{fig:andreev}
\end{figure}
%&&&&&&&&&&&&&&&&&&&&&&&&&&&&&&&&&&&&&&&&&&&&&&&&&&

The structure of ferrons is stabilized by the existence of Andreev states, induced by the
spatial variations of the pairing field $\Delta$ where the majority of spin particles 
are stored, see Fig~\ref{fig:andreev}. It was shown that at the unitarity
the structure of the droplet remains preserved even under dynamic evolution
including stretching and collisions with other droplets~\cite{ferron1}.
In the case of a single ferron, the lowest-energy
condition guarantees that the shape of the ferron remains spherical, and its radius
is a function of polarization. The relation between radius and polarization
reflects the fact that spin excess can be stored in Andreev states, and their number scales
with the radius. Therefore, it is easy to realize that in the case of a spherical three-dimensional (3D) ferron, the
size (radius) $R$ scales as $| N_{\uparrow}-N_{\downarrow} |^{1/2}$, whereas in the case of the 2D
system (or cylindrical ferron), the relation is linear~\cite{ferron2}.
It is also possible to create a spherical ferron with multiple concentric nodal surfaces.
Recently, the ferron like structures have been generated within an extension
of the Ginzburg-Landau (GL) approach, which allows for consideration of the spin-imbalanced system \cite{buzdin_gl}.
Within a certain parameter range of the GL model stable circular solutions in 2D
have been found corresponding to circular ferrons with single or multiple nodal lines. 
They have been described as {\em ring solitons} although their structure coincides with that of ferrons. 
The interaction between ferrons mediated by the superfluid has been determined~\cite{babaev}.

In this paper, we investigate the dynamic properties of a ferron from the
BCS regime towards the unitary point. We
show that the ferron possesses a certain effective mass that scales with its surface.
It is also characterized by a critical velocity that
cannot be exceeded whereas moving through the superfluid environment,  
proportional to the chemical potential difference between the majority and the minority spin components. We discuss the implications of these findings.

\section{\label{sec2}Effective mass}
In the case of nonzero polarization, the nodal line (surface) of the pairing field may acquire
stability as soon as Andreev states become populated. 
Clearly, the nodal line shares the property of the vortex line (the phase changes abruptly by $\pi$) which 
cannot end inside a superfluid. It may either form the closed
structure (e.g., sphere in 3D) or end at the boundary, where the density drops to zero.
Similarly, as in the case of a vortex, one may ask the question: What are the laws 
of dynamics governing the motion of such nodal structures traveling through the superfluid?
In the case of vortices, the answer to these questions gave rise to the formulation 
of the filament model, which accurately predicts dynamics of vortices
and can be applied to describe turbulence phenomenon~\cite{schwarz}.
In order to be able to formulate an effective theory, one needs to extract the inertia of
the object and determine the conservative and dissipative forces present when moving in a
superfluid environment. In this paper, we focus on the effective mass of the ferron.
Although, in general, the determination of the mass of an impurity immersed in a fermionic environment
is a challenging problem~\cite{rosh} due to the presence of the pairing gap 
the problem facilitates considerably.

We determine the mass as the response of the system with ferron, being exposed to the 
superflow characterized by the wave-vector $2 {\bf q}$. Namely, we consider the pairing field
$\Delta({\bf r})$ which in the limit of long-distance $R$ from the ferron behaves as
$\lim_{R\to\infty} \Delta({\bf r}) =  |\Delta| \exp(i 2{\bf q}\cdot{\bf r} + i\phi ) $, with
$\phi$ being an arbitrary overall phase.
However, instead of considering the superflow, we change the reference frame to the one
moving with velocity ${\bf q}$ (we use units: $\hbar = m = 1$). In this case it is sufficient to 
apply the transformation:
$u_{n,\uparrow\downarrow}(\bm{r}) \rightarrow \exp(i{\bf q}\cdot{\bf r})u_{n,\uparrow\downarrow}(\bm{r}),
v_{n,\uparrow\downarrow}(\bm{r}) \rightarrow \exp(-i{\bf q}\cdot{\bf r})v_{n,\uparrow\downarrow}(\bm{r})$
to transform the initial Bogoliubov-de Gennes (BdG) equations (for $u_{n,\uparrow}, v_{n,\downarrow}$  components) 
with the superflow to
\begin{align}\label{eq:hfbspin}
\begin{gathered}
\mathcal{H}({\bf q})
\begin{pmatrix}
u_{n,\uparrow}(\bm{r})\\
v_{n,\downarrow}(\bm{r})
\end{pmatrix}
=E_n
\begin{pmatrix}
u_{n,\uparrow}(\bm{r})\\
v_{n,\downarrow}(\bm{r})
\end{pmatrix},
\end{gathered}
\end{align}
with Hamiltonian, 
\begin{align}\label{eq:hfbspin2}
\begin{gathered}
\mathcal{H}({\bf q})=
\begin{pmatrix}
-\frac{1}{2}(\grad + i{\bf q} )^2 -\mu_{\uparrow}   & \Delta(\bm{r}) \\
\Delta^*(\bm{r}) & \frac{1}{2}(\grad - i{\bf q} )^2 + \mu_{\downarrow},
\end{pmatrix},
\end{gathered}
\end{align}
where $\mu_{\uparrow,\downarrow}$ are chemical potentials for two spin components. The quasi particle wave functions define densities,
 \begin{eqnarray}
 n_{\sigma}(\textbf{r}) &=& \sum_{n} |v_{n,\sigma}(\textbf{r})|^2 f_{\beta}(-E_n), \\
 \tau_{\sigma}(\textbf{r}) &=& \sum_{n} |\nabla v_{n,\sigma}(\textbf{r})|^2 f_{\beta}(-E_n), \\
 \nu(\textbf{r}) &=& \sum_{n}  v_{n,\downarrow}^{*}(\textbf{r}) u_{n,\uparrow}(\textbf{r})\frac{f_{\beta}(-E_n)-f_{\beta}(E_n)}{2}, \\
 \textbf{j}_{\sigma}(\textbf{r}) &=& \sum_{n} \textrm{Im}[v_{n,\sigma}(\textbf{r}) \nabla v_{n,\sigma}^{*}(\textbf{r})]f_{\beta}(-E_n),
 \end{eqnarray}
where $\sigma$ denotes the spin orientation and $f_{\beta}(E_n) = 1/(e^{E_n/T}+1)$ is the Fermi-Dirac distribution. The finite temperature $T$ has been used for numerical convenience with  
$T/T_c \approx 10^{-5}$, where  the critical temperature is calculated from 
well-known BCS result $\Delta/T_c = 1.76$. The pairing field, $\pair{r}$ is calculated self-consistently,
\begin{equation}
\pair{r}= -g_\textrm{eff}\nu(\textbf{r}),
\end{equation}
where $g_\textrm{eff}$ is the coupling constant which is tuned to obtain
the required strength of the pairing field. For more details of 2D calculations see Appendix~\ref{appA}.

The transformation of $u_{n,\uparrow}, v_{n,\downarrow}$ amplitudes to the moving frame induces the
transformation of the currents:
${\bf j}_{\uparrow\downarrow}(\bm{r}) \rightarrow {\bf j}_{\uparrow\downarrow}(\bm{r}) - {\bf q}n_{\uparrow\downarrow}(\bm{r})$,
where the term ${\bf q}n_{\uparrow\downarrow}(\bm{r})$ corresponds to the uniform motion of
spin-up ($n_{\uparrow}$) and spin-down ($n_{\downarrow}$) components, respectively.
Consequently, in this reference frame, the resulting currents represent perturbation to the superflow
induced by the presence of impurity.
As a result we may define the effective mass of the ferron as a static response 
${\cal R}$,
\begin{equation} \label{mass_eff}
M_\mathrm{eff} = \lim_{q\to 0}{\cal R}({\bf q}) = \lim_{q\to 0}
\frac{ | \int d^3 r ({\bf j}_{\uparrow}+  {\bf j}_{\downarrow}) | }{|{\bf q}|}.
\end{equation}

The effective mass contains two components. The first one is related
to the number of majority spin particles that are accumulated inside the ferron
and are dragged through the superfluid by the nodal structure. 
The second component comes from the modification of the surrounding environment by
moving impurity,
\begin{equation}\label{eq:meff}
M_\mathrm{eff} = M_\mathrm{pol} + \delta M = (N_{\uparrow} - N_{\downarrow}) + \delta M.
\end{equation}

Note that the term $M_{\mathrm{pol}}$ scales with the surface since 
$M_{\mathrm{pol}} \propto R^{2}$ ($\propto R$ in 2D) due to the relation between the
polarization and the ferron radius (see Ref. \cite{ferron2}). The second term $\delta M$
scales with the volume of the impurity, but it also depends on the difference between densities
inside and outside the ferron and vanishes when they are equal [see Eqs. (\ref{c5}) and (\ref{c6})]. 
Consequently, in the limit of large ferron radius,
the densities inside and outside become the same, and, therefore, the contribution $M_{\mathrm{pol}}$
becomes dominant. On the contrary, for smaller radii, $\delta M$ may contribute significantly
to the effective mass.
One may also expect that the effective mass is modified with increasing pairing strength ($\Delta/\eF$)
as it corresponds to moving towards the irrotational hydrodynamic limit.

In order to resolve quantitatively these %apparently non obvious 
issues we have performed a series of calculations
in 2D and determined the response function~(\ref{mass_eff}). 
We have applied the BdG approach varying the value of $\Delta/\eF$ from $0.36$ to $0.55$.
Subsequently, we have determined the limit $|{\bf q}| = q\to 0$ numerically
and extracted the effective mass as a function of spin-imbalance 
and pairing gap.
The calculations have been performed in a box with lattice size $70^2$ 
with Fermi momenta for $k_{\textrm{F}\uparrow\downarrow}{}\approx 1$. 
The W-SLDA toolkit has been used for the calculations~\cite{SuppressedSolitonicCascade,PRL__2014,WSLDAToolkit}. 
We have evaluated the total current for a series of velocities $q/v_{\mathrm{F}}= 0.01,~0.02,~0.03,...$ 
until the ferronic configuration is destroyed by 
the currents. For velocities $q\lesssim 0.04$, we found that the linear relation
between the current and the velocity holds with very good accuracy.
We have also analyzed the stability of the results with respect to the size of the box by evaluating
effective mass, in a box with lattice size $100^2$ and found an agreement with accuracy better than $1\%$. 
%&&&&&&&&&&&&&&&&&&&&&&&&&&&&&&&&&&&&&&&&&&&&&&&&&&
\begin{figure}[t]
   \begin{center}
   \includegraphics[width=\columnwidth, trim=0 0 10 0, clip]{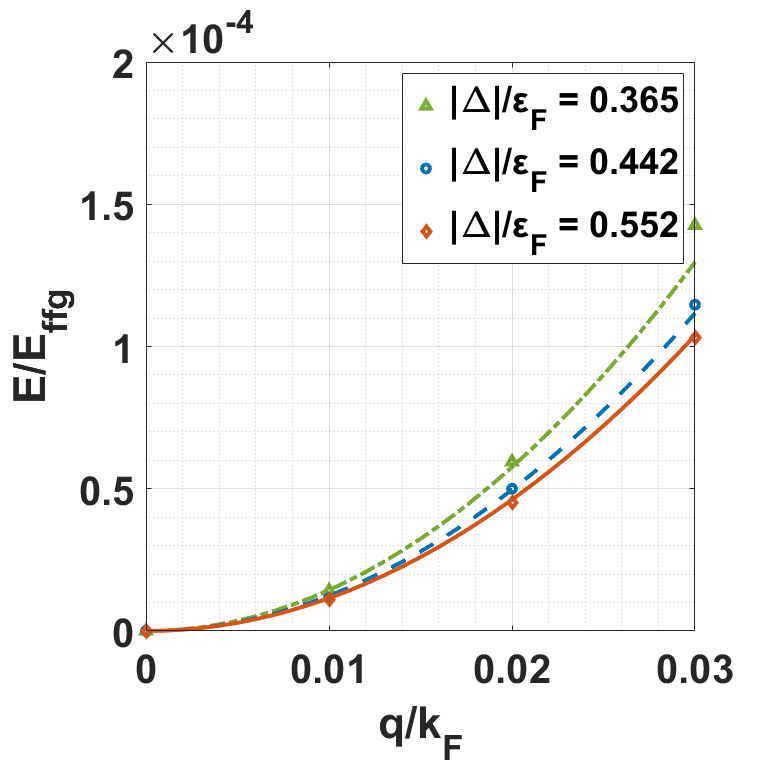} 
   \end{center}\vspace{-3mm}
   \caption{The excitation energy of the system  $E_{\textrm{ex}}(q)$ as a function
   of velocity $q$ obtained in BdG calculations (lines).
   The kinetic energy of the ferron $E_{\textrm{f}}(q)$ obtained using the
   extracted effective mass (points).
   Both energies are shown in units of noninteracting Fermi gas $E_{\textrm{ffg}}$.
   The spin-imbalance in the system is $\delta N = 41$. 
   The lattice size is $(70\kF^{-1})^2$ where $\kF\approx1$.}
   \label{fig:energycomparison}
\end{figure}
%&&&&&&&&&&&&&&&&&&&&&&&&&&&&&&&&&&&&&&&&&&&&&&&&&&
Finally, we have checked that extracted effective mass when plugged into equation
$E_{\textrm{f}}(q) = \frac{1}{2}M_{\textrm{eff}}q^2$,
reproduces reasonably well the behavior of the energy change (computed as the volume integral of the BdG functional)
$E_{\textrm{ex}}(q) = E(q)-E(0)$. Therefore, $E_{\textrm{ex}}(q)$ gives the contribution to the energy coming from the ferron's response to the superflow. In \fig{fig:energycomparison} we compare these values to the kinetic energy of the ferron moving in a superfluid environment $E_{\textrm{f}}(q)$, where $M_{\textrm{eff}}$ is extracted by means of Eq.~(\ref{mass_eff}). For low-$q$ values, good agreement between two approaches is obtained. For additional technical details of the effective-mass extraction procedure see Appendix~\ref{appB}. 

\begin{figure}[t]
   \begin{center}
   \includegraphics[width=\columnwidth]{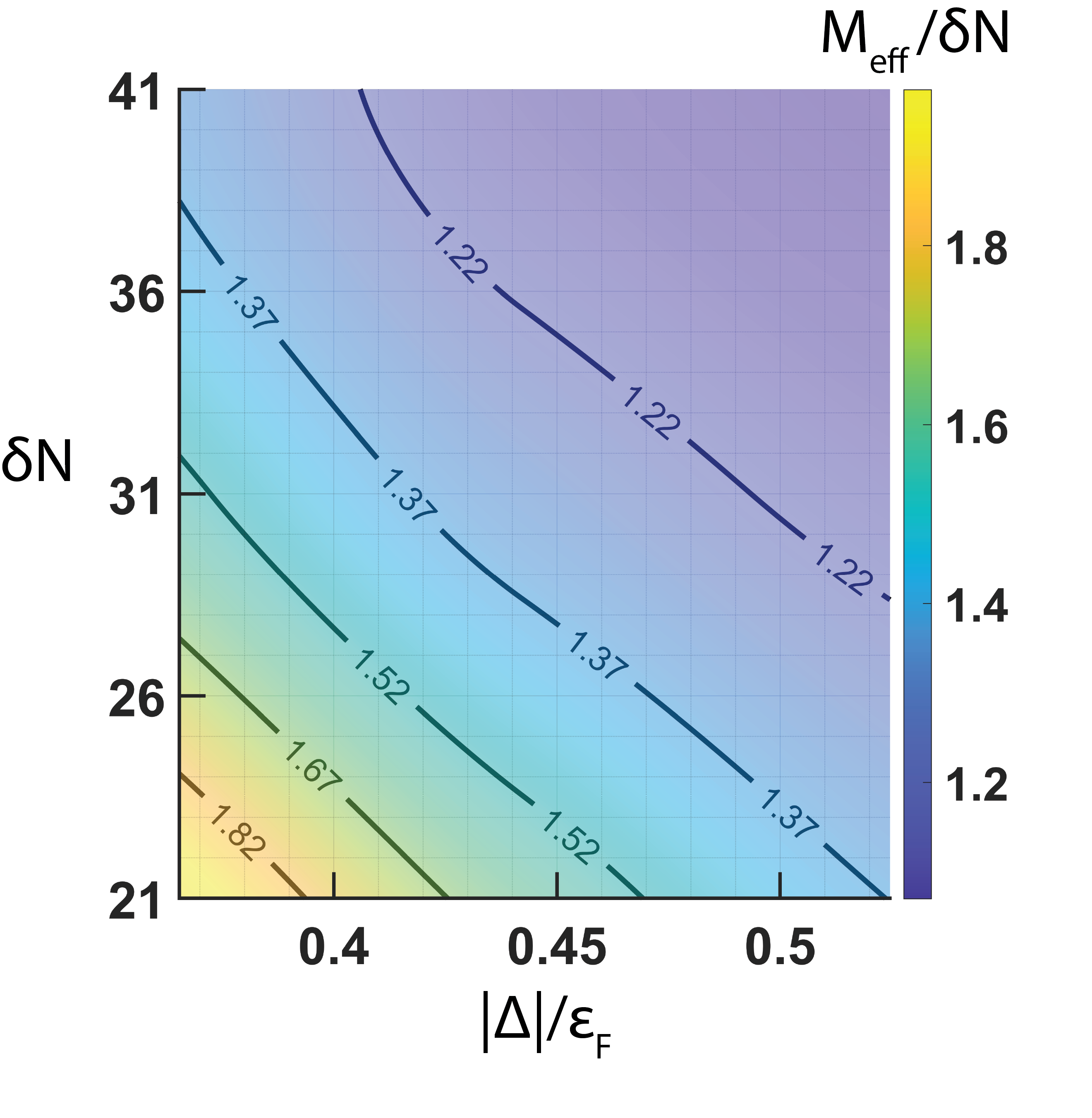} 
   \end{center}\vspace{-3mm}
   \caption{The effective mass $M_\mathrm{eff}$ as a function of the magnitude of the pairing field $\abspair{}/\eF$ and the spin-imbalance $\delta N = N_{\uparrow} - N_{\downarrow}$. In all cases the total number of particles in the simulation box is $N = N_\uparrow + N_\downarrow = 770$, and the Fermi momentum is $\kF=\sqrt{2\eF} \approx 1$. The values of $\delta N = 21, 41$ correspond to the ferron radii 
$R\approx 4.0\xi$ and $R\approx8.3\xi$, respectively, where $\xi = \frac{1}{\kF} \frac{\eF}{|\Delta|}$.}
   \label{fig:effective_mass}
\end{figure}

The results, for the effective mass, shown in the Fig. \ref{fig:effective_mass} indicate that the contribution
coming from the flow which is induced in the superfluid medium $\delta M$ is a correction to 
the dominating term $M_{\textrm{pol}}$, except for the small ferron size of the order of
coherence length. In order to understand this result one may notice that
in pure irrotational hydrodynamics in 2D the contribution to $\delta M \propto S \frac{n_{out}-n_{in}}{n_{out}+n_{in}}$, (where $n_{in}$, $n_{out}$ correspond to superfluid
density inside and outside impurity, respectively, and $S$ is its area)
and, thus, it vanishes if $n_{in}\rightarrow n_{out}$. For more details on irrotational hydrodynamics, see Appendix~\ref{appC}.
%&&&&&&&&&&&&&&&&&&&&&&&&&&&&&&&&&&&&&&&&&&&&&&&&&&
%&&&&&&&&&&&&&&&&&&&&&&&&&&&&&&&&&&&&&&&&&&&&&&&&&&
Clearly, the largest discrepancy between the magnitude of
the pairing field inside from its bulk value occurs for small ferrons (and weak pairing). 
In that case due to the fact that coherence length is on the order of the ferron size (i.e., $\xi \approx R$) the value of the pairing
gap inside is smaller than outside. It also implies that the polarization inside the ferron does not vanish 
completely.
As a consequence, there is a larger contribution coming from the flow than in the case
of large ferron. In the latter case, the magnitude of the pairing field inside the ferron is the same 
as outside and therefore, the perturbation related to the flow occurs effectively around
the pairing nodal area.

\section{\label{sec3}Critical Velocity}
While moving through a superfluid the structure of the ferron is affected. The spherical ferron
has a characteristic spectrum of Andreev states. It consists of almost degenerate states at $E_{\pm} \approx \pm\frac{1}{2}\delta\mu $,
where $\delta\mu=\mu_{\uparrow}-\mu_{\downarrow}$
(see Appendix~\ref{appD}).
In the case of a small ferron the degeneracy is lifted due to the tunneling effect through the interior of the ferron, %$\delta E(l)$, 
which, however, decreases exponentially with its size~\cite{ferron3}.
Due to circular (in 2D) or spherical (in 3D) symmetries of the ferron, the states
can be labeled by quantum numbers associated with angular momenta. Namely,
in the 2D case the magnetic quantum number 
$m = \langle \hat{L}_{z} \rangle/\int d^{2}r |v_{i}({\bf r)}|^2$ ($i=\uparrow,\downarrow$) can be used to label states (the $z$ axis is perpendicular
to the plane on which ferron resides). 
The spectrum of these states correspond to the range:
$-\kF R\lesssim m \lesssim \kF R$, where $R$ is the ferron radius.
The 3D ferron possesses the same structure of Andreev states with an additional $2l+1$ degeneracy of each state labeled by the orbital quantum number associated with the $\hat{L}^{2}$ operator.
Apart from these degenerate states which accumulate the spin polarization, there 
is a small fraction of
states with $m \approx \pm\kF R$, which energy varies with angular momentum.
These states can be interpreted as related to periodic orbits located in the nodal region representing trajectories between pairing potential of the same phase~\cite{ferron3}.

\begin{figure}[t]
   \begin{center}
   \includegraphics[width=\columnwidth, trim=50 0 50 0, clip]{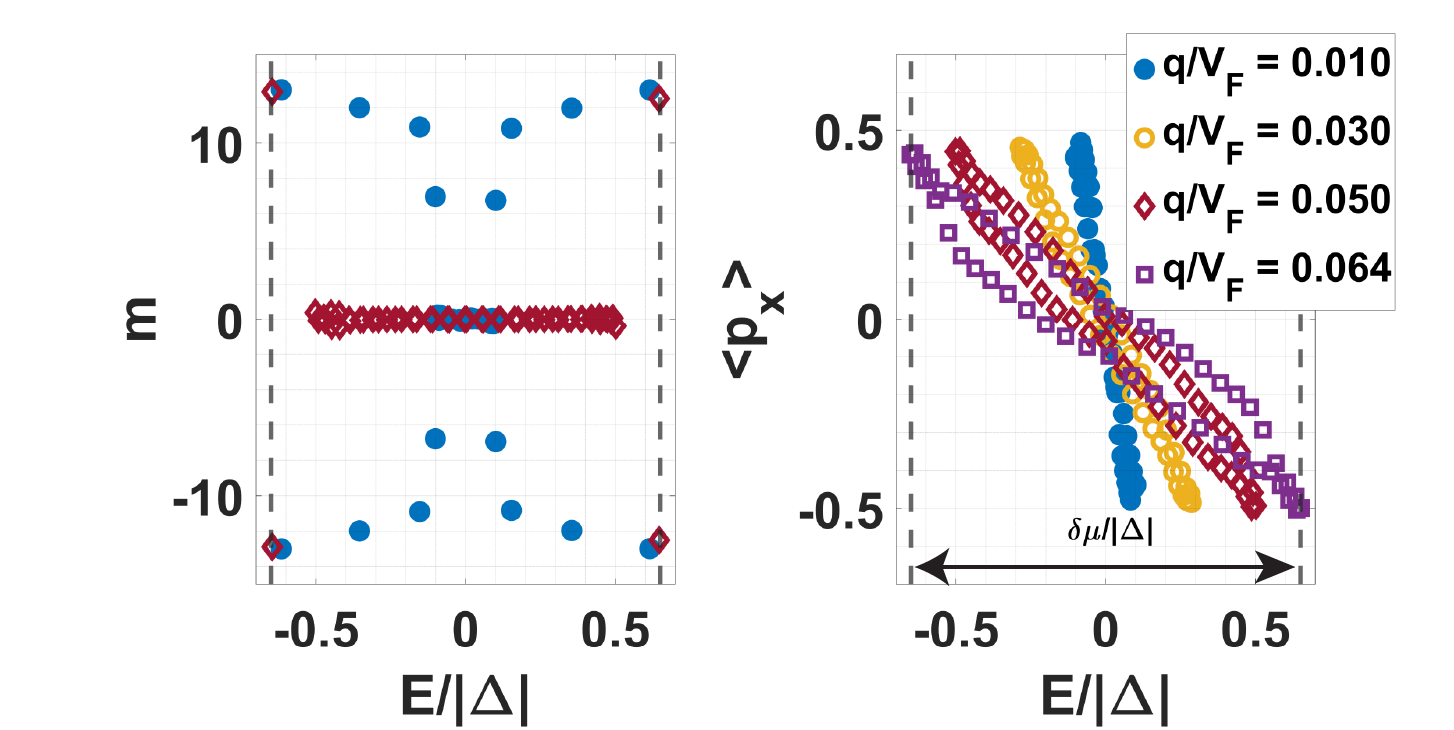} 
   \end{center}\vspace{-3mm}
   \caption{Structure of the spectrum of Andreev states exposed to different strengths of superflow. {\bf Left subfigure:} magnetic quantum numbers $m$ ($m = \langle L_z \rangle /|v|^2$, where $|v|^2$ denotes occupation probability of the state) corresponding to Andreev states are shown
for two velocities of the superflow: $q/v_{\mathrm{F}}=0.01$ (filled circles) and $q/v_{\mathrm{F}}=0.05$ (empty diamonds), where $v_{\mathrm{F}}$ denotes Fermi velocity.
{\bf Right subfigure:} the expectation value of the momentum operator component, parallel to the direction 
of the superflow is shown for Andreev states. The quasiparticle energies have been shifted by 
$\frac{1}{2}\delta \mu$ and therefore the plots on both subfigures
possess symmetry with respect to $E=0$. The shifted energy values
corresponding to $\pm \frac{1}{2}\delta \mu$ have been denoted by vertical dashed lines.
The spin imbalance corresponds to $\delta N = 31$ ($ R\approx6.2\xi $) and the strength of the pairing field 
$\abspair{}/\eF = 0.44$.}
   \label{fig:andreev_states}
\end{figure}

It is important to realize that the stability of the ferron is exclusively related to the structure of Andreev states. 
When the ferron is moving through the superfluid or, equivalently, when it is exposed 
to the superflow, the structure and energies of these states are modified.
The perturbation is induced by the pairing field, which is affected 
by the superflow. In particular, the phase of the pairing field is modified 
on both sides of the nodal line, depending on its orientation with respect to the direction of superflow. Namely, the spherically symmetric pairing field becomes perturbed by the superflow in the following way:
$\Delta_{0} (r) \rightarrow \Delta ({\bf r})=
\tilde{\Delta}_{0} ({\bf r})\exp( 2 i{\bf q}\cdot{\bf r} )$.
Neglecting in the first approximation the modification of the magnitude of the initial pairing field associated with the ferron, i.e., 
$\tilde{\Delta}_{0} ({\bf r})\approx \Delta_{0} (r)$, it is easy to show that
energies of Andreev states forming degenerate branches $E_{\pm}=\pm\frac{1}{2}\delta\mu$
will be splitted 
proportionally to $q$ (see Appendix \ref{appD}).
The modification of the spectrum of states inside the ferron can be seen in 
Fig. \ref{fig:andreev_states}.

%&&&&&&&&&&&&&&&&&&&&&&&&&&&&&&&&&&&&&&&&&&&&&&&&&&
%&&&&&&&&&&&&&&&&&&&&&&&&&&&&&&&&&&&&&&&&&&&&&&&&&&

All Andreev states inside the ferron at rest have the vanishing expectation value 
of linear momentum.
When the ferron is moving, they acquire a non zero component of momentum in the direction
of the flow. The most affected states are those with small angular momenta.
As the velocity increases, more states become affected, contributing to
the splitting width. Eventually, at a certain superflow velocity, the
splitting width becomes equal to $\delta\mu$ and, consequently, the lowest positive energy
Andreev state reaches zero energy.
This can be seen in Fig.\ref{fig:andreev_states}, where in the right panel
the spectrum of states is plotted for various superflow velocities.
At the critical velocity, the spectrum of states reaches zero energy, and quasiparticle excitations lead to ferron instability and subsequent decay.
Consequently, one may attribute to each ferron a certain critical velocity $\vcrit$
which constitute its maximum velocity when moving through the uniform superfluid.
Since the splitting width of Andreev states is proportional to the superflow
(see Appendix \ref{appD}) one may conclude
that critical velocity is proportional
to the chemical potential difference between the majority and the minority spin
components $\delta\mu$. 

\begin{figure}[t]
   \begin{center}
   \includegraphics[width=\columnwidth]{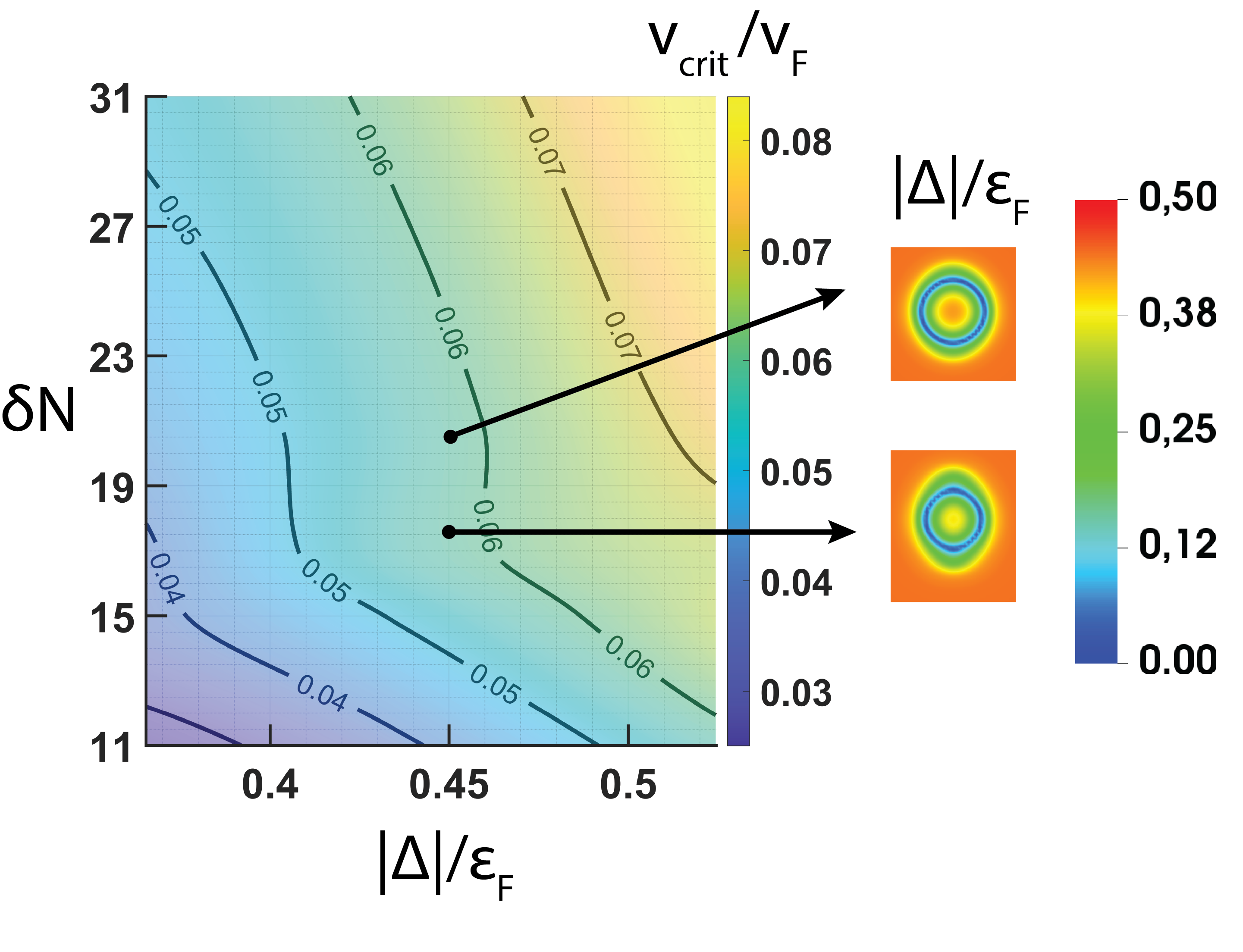} 
   \end{center}\vspace{-3mm}
   \caption{The ferron critical velocity as a function of the magnitude of 
   the pairing field $\abspair{}/\eF$ and the spin-imbalance $\delta N$. The simulation settings are the same as for Fig.~\ref{fig:effective_mass}. The inset shows an example of two different sizes of ferrons having the same critical velocity where the smaller ferron is deformed.}
   \label{fig:velocity_contour}
\end{figure}

In order to validate the above statement and to make a quantitative estimation
of $\vcrit$ we have performed a series of numerical calculations in 2D.
In Fig. \ref{fig:velocity_contour}
the critical velocity in units of Fermi velocity has been shown as a function of pairing gap and ferron size. It is of no surprise that the larger sizes of ferrons admit larger velocities.
Clearly, it is related to the fact that that they require larger spin
polarization and, consequently, larger chemical potential difference. 
As a consequence, ferrons with larger polarizations can move with higher velocities through the medium.
The relation between critical velocity and polarization that turn out to be approximately linear in 2D as expected (apart from deviations induced by deformation changes at the vicinity of critical
velocities) represent an interesting manifestation of the relation between 
spatial pairing field 
modulation and its dynamic properties. In 3D, all the arguments remain valid, however, one may expect
that due to additional degeneracy, the relation between critical velocity and polarization
will read $\vcrit \propto \sqrt{\delta N}$.
The deviations which are visible in Fig.~\ref{fig:velocity_contour} are attributed to
the {\em shell effects} related to Andreev states. 
Namely, for velocities close to the $\vcrit$, some ferrons become deformed, which can be
seen in the inset in Fig. \ref{fig:velocity_contour} 

%&&&&&&&&&&&&&&&&&&&&&&&&&&&&&&&&&&&&&&&&&&&&&&&&&&
%&&&&&&&&&&&&&&&&&&&&&&&&&&&&&&&&&&&&&&&&&&&&&&&&&&

\section{\label{sec4}Induced motion of ferron and interaction with a vortex}
\begin{figure}[t]
   \begin{center}
   \includegraphics[width=\columnwidth]{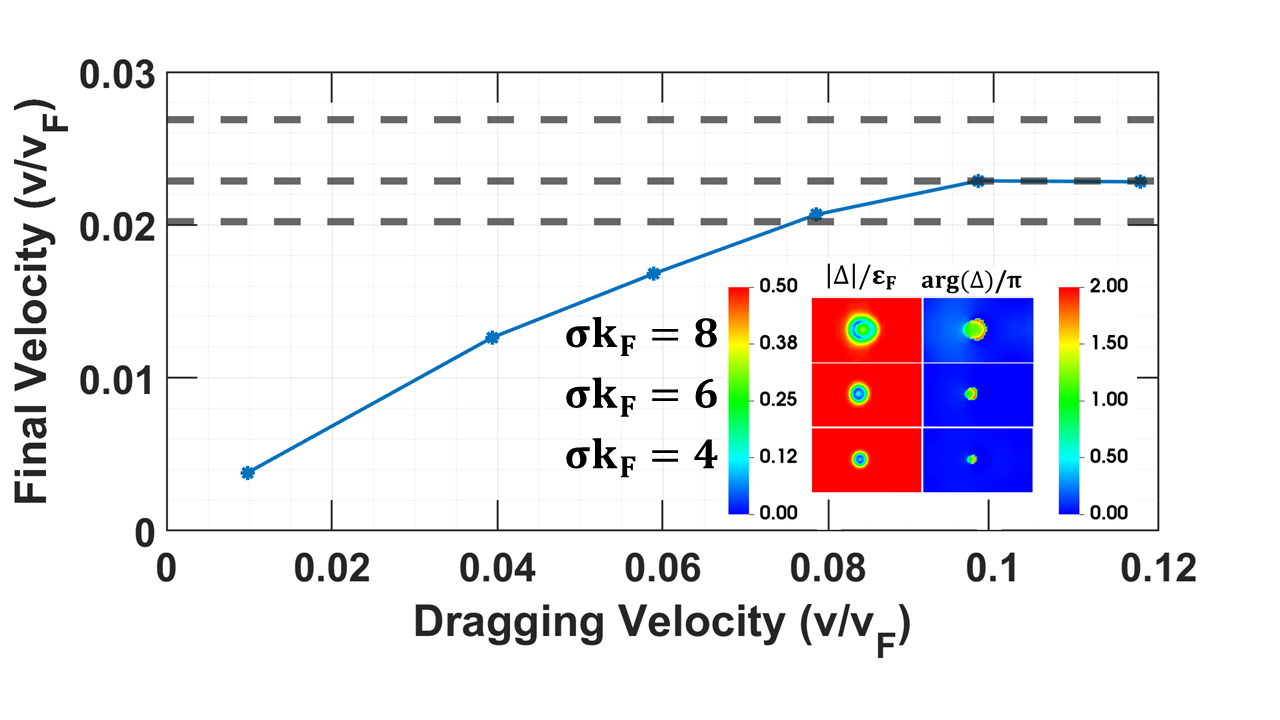} 
   \end{center}\vspace{-3mm}
   \caption{Velocity of the ferron in the final state as a function of the dragging velocity. The time-dependent spin-selective potential is dragged along the $x$-axis during its application. The horizontal dashed lines shows the "plateau" of the final velocity for various sizes of ferrons corresponding to: $\sigma\kF = 8$, $\sigma\kF = 6$, $\sigma\kF = 4$ from top to bottom, where $\sigma$ is the width of the Gaussian potential. The inset shows the absolute value of the pairing field in the left column, whereas in the right column, the phase of the pairing field is shown. The images are taken after the external potential is turned off whereas the ferron is moving. 
%as a result of the initial momentum generated by the external potential. 
All three configurations correspond to $\vdrag/v_{\mathrm{F}} = 0.06$. In the simulations we used a box of size $53\xi \times 31\xi \times 31\xi$ in the $x$,$y$,$z$ dimensions, respectively. Fermi momentum $\kF \approx 1$.
For full movies see the Supplemental Material~\cite{supplemental}.}
   \label{fig:velocity_dynamics}
\end{figure}

The results presented in the previous section can also be looked at from another perspective.
Namely, assume that one creates a ferron as an excited configuration in an unpolarized superfluid 
medium. This can be achieved by dynamically applying a spin-selective potential, which
will locally break Cooper pairs. If the potential is applied for a sufficiently long time,
it allows the pairing field to adjust by developing a nodal surface. 
It was shown in Ref. \cite{ferron1} that such configuration is stable despite the fact 
that the ferron is surrounded by phonon excitations.
Taking into account results from the previous section, one may ask the following question:
What is going to happen if one attempts to accelerate ferron beyond the critical velocity?
In order to investigate this issue, we have performed the following time-dependent simulations in 3D.
We have applied a spin-selective potential in the form of the Gaussian, by following the procedure described in~\cite{ferron1}. The procedure is based on the application of time-dependent potential of the form:
\begin{equation}
V_{s}(\bm{r},t)=\lambda_{s} A(t) \exp\left[  -\frac{(x_0 + \vdrag t)^2 + y^2 + z^2}{2\sigma^2} \right].
\label{vt}
\end{equation}
%\todo{BT: check if formula is modified correctly: changed $v$ label} 
This potential is repulsive for spin-up components, $\lambda_\uparrow = +1$, and attractive for spin-down componentsf $\lambda_\downarrow = -1$. The width of the Gaussian potential $\sigma$ sets the size of the ferron. The amplitude $A(t)$ is a time-dependent function that starts as $0$ and is slowly increased to its maximum value, and then it is decreased back to $0$. Details of the implementation of the spin-selective potential are provided in Appendix~\ref{appE}. When the ferron is created,
we have accelerated the potential, which was dragging the ferron through the superfluid
with velocity $\vdrag$. 
Subsequently, we have removed the potential allowing the ferron to move freely.
It has been found that the ferron, after switching off the potential, continues
its motion although it always slows down to the velocity $\vfin$ (see \fig{fig:velocity_dynamics}). Still, for velocities $\vdrag \ll \vcrit$, the
relation between $\vdrag$ and $\vfin$ is approximately linear. 
However, when $\vdrag$ becomes large enough,
the velocity $\vfin$ saturates and attempts to increase the ferron velocity beyond
a certain value fail. Note that the results shown in \fig{fig:velocity_dynamics} for different sizes of the ferron are consistent with the static results; the critical velocity increases 
with the size of the ferron. 

%&&&&&&&&&&&&&&&&&&&&&&&&&&&&&&&&&&&&&&&&&&&&&&&&&&
%&&&&&&&&&&&&&&&&&&&&&&&&&&&&&&&&&&&&&&&&&&&&&&&&&&

\begin{figure}[t]
   \begin{center}
   \includegraphics[width=\columnwidth]{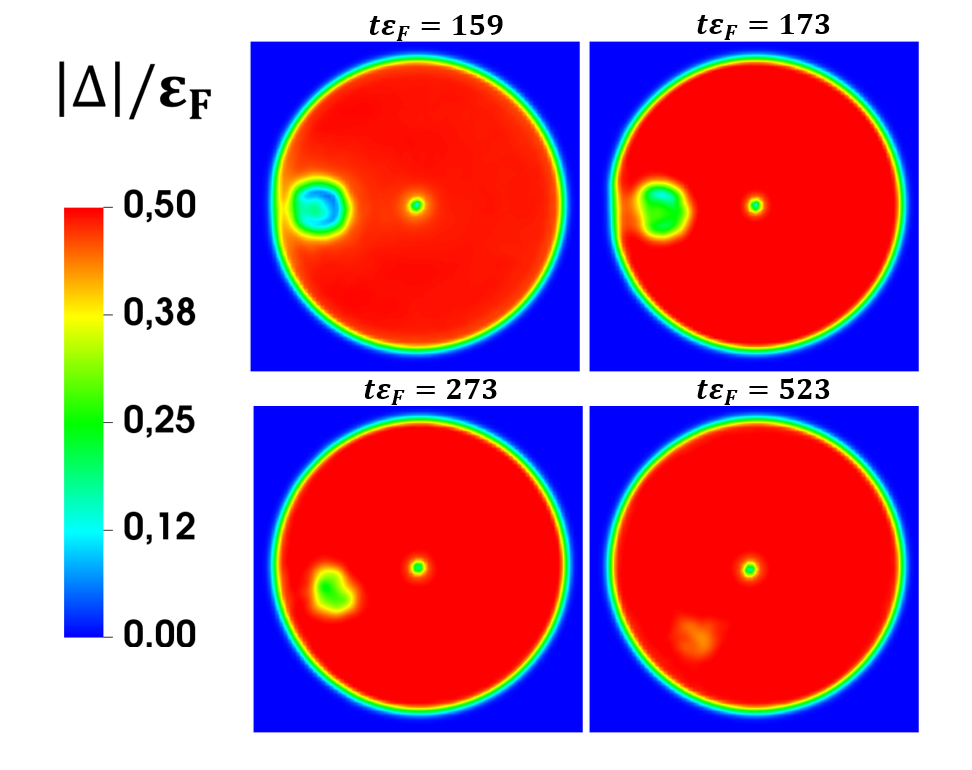} 
   \end{center}\vspace{-3mm}
   \caption{Snapshots showing the attempt to create a stable ferron solution in the presence
   of the vortex. The time-dependent potential to generate the ferron is turned off at $t\eF=150$. The vortex, with the core located in the center, creates currents rotating counter-clockwise. It is visible that the ferron is destroyed because of these currents. The polarization inside the ferron is pushed to the boundary of the system. For detailed information and full movies see the Supplemental Material \cite{supplemental}.}
   \label{fig:vortex_ferron}
\end{figure}

The existence of the critical velocity has yet another important consequence when it comes to the possibility of creating vortices in the system with ferrons. Namely, it is possible to have a coexistence
of a vortex and a spherical ferron as long as the distance between the vortex core and the ferron
is large enough. In this case, the superflow generated by a vortex is weak enough to support 
the existence of the ferron solution. 
On the other hand, an attempt to create a ferron in the vicinity of the vortex core fails
which is shown in the Fig.~\ref{fig:vortex_ferron}. In this particular simulation we generated the ferron of radius $rk_{\mathrm{F}}=6$ (by the spin-selective Gaussian potential) in the distance $d\kF=24$ from the core. At the point where the ferron is closest to the vortex, the induced velocity $v = \frac{\hbar}{2mr}\approx 0.028 v_{\mathrm{F}}$ is higher than the critical velocity for this case $\vcrit\approx 0.024v_{\mathrm{F}}$. Therefore, the snapshots reveal stages of ferron decay. For more details on simulation, see Appendix~\ref{appE}.
One expects that large ferrons which are characterized by higher critical velocity may
be created closer to the vortex core. However, in this case, effects related to non uniformity
of the superflow within the volume of the ferron may become important.

%&&&&&&&&&&&&&&&&&&&&&&&&&&&&&&&&&&&&&&&&&&&&&&&&&&
%&&&&&&&&&&&&&&&&&&&&&&&&&&&&&&&&&&&&&&&&&&&&&&&&&&

\section{\label{sec5}Conclusions}
We have investigated the dynamical properties of ferrons related to their motion through
the superfluid. We have extracted the effective mass of this object which turned out
to be related mainly to spin imbalance with a small correction coming from induced superfluid flow.
Only for small ferrons (of sizes on the order of a few coherence lengths), the latter
contribution becomes important. It implies that the effective mass
scales rather with the surface than the volume of impurity.
We have also shown that each ferron is characterized by a certain critical velocity that
cannot be exceeded whereas moving through the superfluid environment. The critical velocity is proportional to the chemical potential difference
between the majority and the minority spin components, and consequently, it
increases with the ferron size. It was demonstrated that it is not possible to
accelerate the ferron dynamically by dragging it beyond a certain velocity. For the same reason,
it is not possible to create a stable configuration of the ferron in the vicinity of 
the vortex core.

\begin{acknowledgments}
P.M. would like to thank Centre for Computational Sciences at the University of Tsukuba,
where part of this work has been performed for hospitality.
This work was supported by the Polish National Science Center (NCN) under
Contracts No. UMO-2016/23/B/ST2/01789 (P.M. and B.T.) and No. UMO-2017/26/E/ST3/00428 (GW).
We acknowledge PRACE for awarding us access to resource Piz Daint
based in Switzerland at the Swiss National Supercomputing Centre (CSCS), Decision No. 2019215113. 
We also acknowledge the Global Scientific Information and Computing Center, Tokyo Institute of Technology for resources at TSUBAME3.0 (Project No: hp200115) and the Interdisciplinary Centre for Mathematical and Computational Modelling (ICM) of Warsaw University for computing resources at Okeanos (Grant No. GA83-9).
The contribution of each of the authors has been significant and
the order of the names is alphabetical.
\end{acknowledgments}

\appendix
\section{Details of static BdG calculations in 2D systems} \label{appA}

The total energy density of the system in the BdG approach is expressed through kinetic and anomalous densities:
\begin{equation}\label{funcbdg}
\mathcal{E}_{\small{BdG}} = \frac{\tau_{\uparrow}+\tau_{\downarrow}}{2} + g_\textrm{eff}\nu^{\dagger}\nu.
\end{equation}
We obtain the stationary configuration by minimizing the following functional:
\begin{equation}
F = E-\sum_{s=\{\uparrow,\downarrow\}}\mu_{s}
N_{s}
-\sum_{s=\{\uparrow,\downarrow\}} \int \bm{q}\cdot\bm{j}_{s}({\bf r}) d{\bf r},
\end{equation}
where  $N_{s}=\int n_{s}({\bf r}) d\bf r$ denotes the particle
number of the spin-$s$ component, $\mu_{s}$'s are corresponding chemical potentials, and $E=\int \mathcal{E}_{\small{BdG}}(\bm{r})d\bm{r}$ is the energy. The last term generates the flow in directions given by $\bm{q}$. Minimization of the $F$ functional provides Eqs. (1) and (2) from the main paper. In calculations we used velocity $\bm{q}$ directed along the $x$ direction.  

The ferronic solution corresponds to a particular choice of pairing field $\pair{r}$ which involves a closed nodal line. To capture the ferron 
geometry we imposed the constraint on the pairing potential to have the form:

\begin{equation}
\Delta(\bm{r})=\left\lbrace 
\begin{array}{ll}
-\Delta, & r<R_\mathrm{in},\\
\phantom{-}\Delta, & r>R_\mathrm{out},
\end{array}
\right. 
\end{equation}
To get the ground state of the ferron, we applied the above constraint to the system for a couple of iterations 
during the energy minimization and, subsequently, released it. Values of  $R_\mathrm{in}$ and $R_\mathrm{out}$ are selected in such a way that after convergence, the radius of the ferron is between these values. 
Consequently, the initially imprinted pairing potential captures the main features 
of the ferron, which consist of
outer and inner areas where the phase of the pairing field varies by $\pi$ and
the nodal region of the size of the coherence length where the pairing field vanishes.

The Andreev states inside the circular ferron (at $q=0$) can be labeled by 
eigenvalues of the angular momentum operator component perpendicular to its area (which we
denote by $\hat{L}_{z}$). 
However, due to the degeneracy of states corresponding to positive and negative
eigenvalues of $\hat{L}_{z}$ these states are mixed in numerical calculations and 
do not have well-defined $L_{z}$ values.
Therefore in order to remove this degeneracy we add a small perturbation to the system of Equations~(2) in the form: 
$-\frac{1}{2}( \grad + i{\bf q} )^2 - \tilde{\mu} - \omega {L}_{z}$ where $\omega$ is the radial frequency. We typically set this value to $\omega \approx 0.01\eF$. The perturbation is added only to extract and 
visualize the Andreev states (see Fig.~3 in the paper) and is not applied 
to get the self-consistent solution.

For 2D static calculations, we use a simulation box with a lattice size of $70\kFi$ in the $x$ and $y$ directions. 
We set the Fermi momentum $\kF=\sqrt{2\pi(n_\uparrow + n_\downarrow)} \approx 1$. 

\section{Extraction of the effective mass}\label{appB}
\begin{figure}[t]
   \begin{center}
   \includegraphics[width=\columnwidth, trim=10 40 10 40, clip]{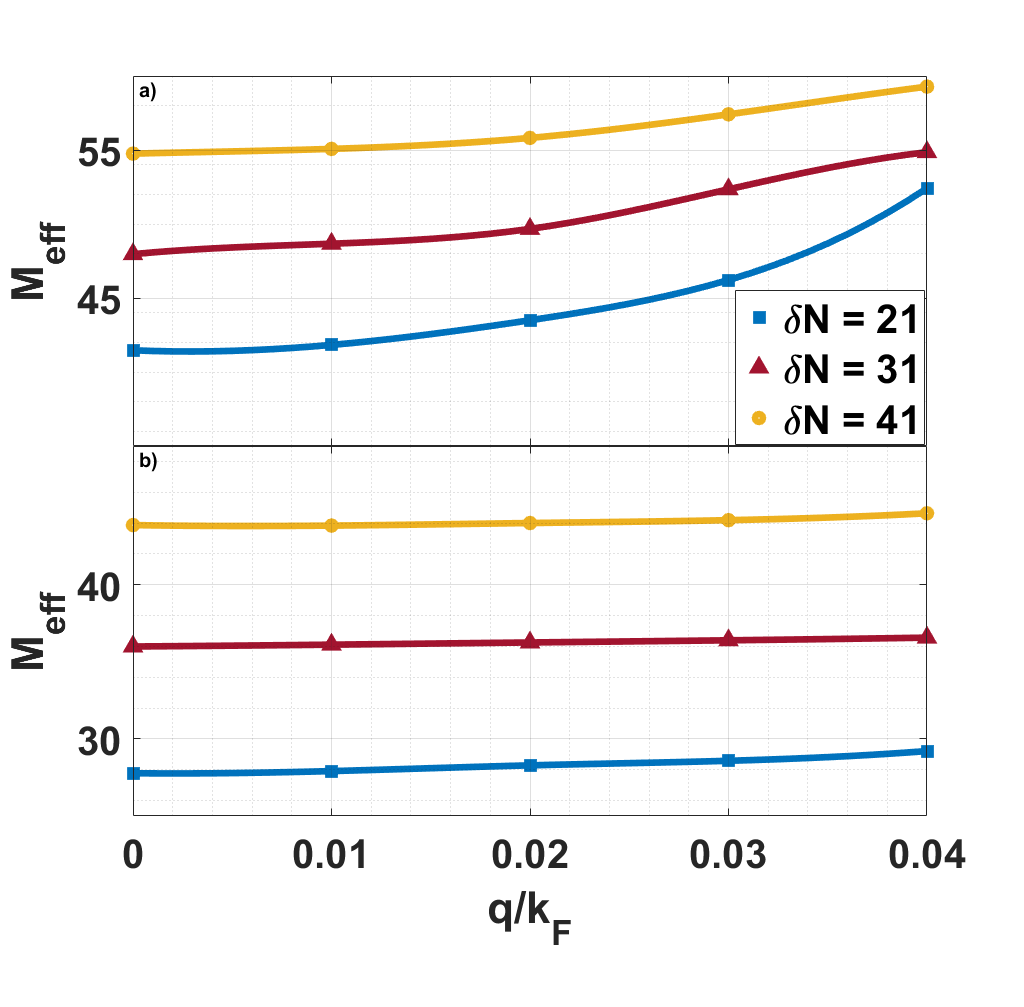} 
   \end{center}\vspace{-3mm}
   \caption{The response function [see Eq.~\ref{mass_eff} in the main text] as a 
   function of the superflow velocity $q$. Symbols correspond to numerical
calculations. Lines are obtained as a result of interpolation. 
The value of effective mass is extracted in the limit of $q\rightarrow 0$ and
denoted by symbols. The results for two values of the pairing field are shown: 
$\Delta/\eF = 0.365$ [panel (a)] and for $\Delta/\eF = 0.552$  [panel (b)]. 
The lattice size is $(70\kF^{-1})^2$ where $\kF\approx1$.}
   \label{fig:extrapolation}
\end{figure}
We introduce a superflow corresponding to the velocity of $q/v_{\mathrm{F}}=0.01,0.02,0.03$ and $0.04$ and calculate the momentum to velocity ratio
by executing the formula~(9) from the main text.
Next, we extrapolate
results to the $q\rightarrow 0$ limit using the cubic Hermite interpolation. In \fig{fig:extrapolation} we show two examples for different pairing strengths. 
As shown in the main text, a ferron in a system with weaker pairing strength has lower critical velocity. Consequently in panel (a) of \fig{fig:extrapolation}  
the response function exhibit more pronounced dependence on $q$ than in
the strong pairing limit shown in panel (b). This is due to the
fact that in the former case the critical velocity is lower and  
the shape of the ferron becomes affected already at relatively small-$q$ values.

\section{Mass of circular impurity in 2D irrotational hydrodynamics}\label{appC}

In this appendix, we present a derivation of the effective mass of circular impurity
that can be obtained in irrotational hydrodynamics. 
Let us consider an impurity of radius $R$ moving with velocity ${\bf v}$ through the superfluid characterized
by the velocity potential $\Phi$,
\begin{equation} \label{pot}
\grad^{2}\Phi({\bf r}) = 
\left ( \frac{\partial^{2}}{\partial x^{2}}+\frac{\partial^{2}}{\partial y^{2}} \right )\Phi({\bf r}) =0.
\end{equation}
Inside the impurity the density is denoted by $n_{in}$ whereas outside by $n_{out}$.
Conditions for the velocity potential at infinity and at the boundary of impurity lead to
\begin{eqnarray}
& &\lim_{r \to \infty}\Phi({\bf r}) = 0 \nonumber \\
& &\Phi({\bf r})|_{r=R^{-}} = \Phi({\bf r})|_{r=R^{+}} \\
& &n_{in}\frac{\partial \Phi}{\partial r}|_{r=R^{-}} - n_{out}\frac{\partial \Phi}{\partial r}|_{r=R^{+}} 
= ( n_{in}-n_{out} ){\bf v}\cdot{\bf n}, \nonumber
\end{eqnarray}
where the last equation is a consequence of the continuity relation for the fluid and ${\bf n}$
denotes the unit vector, normal (outward) to the boundary.
The solutions of eq. (\ref{pot}) reads
\begin{eqnarray}
& &\Phi_{in}({\bf r}) = \frac{n_{in}-n_{out}}{n_{in}+n_{out}}{\bf v}\cdot {\bf r}  \\
& &\Phi_{out}({\bf r}) = \frac{n_{in}-n_{out}}{n_{in}+n_{out}}\frac{R^{2}}{r^{2}}{\bf v}\cdot {\bf r} .
\end{eqnarray}
One can now evaluate the energy of the system, which is stored in the flow,
\begin{eqnarray} \label{c5}
E &=& \frac{1}{2} \int_{r<R} n_{in}\left (\grad\Phi_{in} \right )^2 d^{2}{\bf r}
       \nonumber \\
 &+& \frac{1}{2}\int_{r>R} n_{out}\left (\grad\Phi_{out} \right )^2 d^{2}{\bf r}  = \nonumber \\
&=&\frac{1}{2}\pi R^{2} \frac{(n_{in}-n_{out})^{2}}{n_{in}+n_{out}}v^{2}.        
\end{eqnarray}
From this relation it is clear that one can associate the effective mass of the impurity
with the expression $M^{E}_{\mathrm{eff}}=\pi R^{2} \frac{(n_{in}-n_{out})^{2}}{n_{in}+n_{out}}$.

Another way to extract the effective mass is to evaluate the component of the momentum 
of moving fluid in the direction of ${\bf v}$,
\begin{eqnarray} \label{c6}
\frac{{\bf p}\cdot{\bf v}}{v} &=& 
\int_{r<R} n_{in} \left ( \grad\Phi_{in}\cdot\frac{\bf v}{v} \right ) d^{2}{\bf r} \nonumber \\
&+& \int_{r>R} n_{out} \left ( \grad\Phi_{out}\cdot\frac{\bf v}{v} \right ) d^{2}{\bf r} = \nonumber \\
&=&\pi R^{2} \frac{n_{in}-n_{out}}{n_{in}+n_{out}}n_{in} v.
\end{eqnarray}
The above expression allows to extract the effective mass 
$M^{p}_{\mathrm{eff}}=\pi R^{2} \frac{n_{in}-n_{out}}{n_{in}+n_{out}}n_{in}$
which differs from $M^{E}_{\mathrm{eff}}$ . Differences are due to the fact that
in $M^{p}_{\mathrm{eff}}$ only the component of the current parallel to velocity ${\bf v}$ was taken into account.
Note, however that both contributions are proportional to the area of impurity and both
disappear when $n_{in}\rightarrow n_{out}$.

\section{Andreev states in the presence of superflow} \label{appD}

We consider the impact of superflow on Andreev states inside the ferron.
To capture the ferron geometry we use the schematic potential of the form:
\begin{equation}
\pair{r} = \Delta \left [ \theta(r - R_{out}) - \theta(R_{in} - r) \right ]\exp( 2 i{\bf q}\cdot{\bf r} ),
\label{imprint}
\end{equation}
where $\Delta$ is real and positive and
$\theta$ denotes the Heaviside step function.
The pairing potential reflects the main features of the ferron, which consists of
outer and inner areas of radii $R_{\textrm{out}}$ and $R_{\textrm{in}}$, respectively.
The phase of the pairing field varies by $\pi$ between inner and outer
regions in the absence of superflow. The nodal region
[where $\pair{r}=0$] is of the size of the coherence length $\xi$. 
In the presence of the superflow, the phase pattern is modified by the factor $\exp( 2 i{\bf q}\cdot{\bf r} )$, where ${\bf q}$ defines the direction and magnitude
of the superflow.

According to the Andreev approximation, one decomposes amplitudes $u$ and
$v$ by separating the fast oscillation at the length scale of the Fermi wavelength
and slow variations related to $\xi$. In the case of spin imbalanced system this 
prescription works as long as the difference between Fermi spheres 
of the majority and minority components
is not too large. Providing that this is the case one may associate 
fast oscillations with $\kF$, which is related to the average of chemical potentials
$ \frac{1}{2}\kF^2 = \frac{1}{2}(\mu_{\uparrow} + \mu_{\downarrow}) $ and
BdG equations can be reduced to first-order differential equations,
\beq
\tbtham{-i{\bf \kF}\cdot\grad}{\pair{r}}{\pair{r}^{*}}{i{\bf \kF}\cdot\grad}
{\left( \begin{array}{cc}
		\utp{r} \\ \vtm{r}
	\end{array} \right)} = E^{(+)}
{\left( \begin{array}{cc}
		\utp{r} \\ \vtm{r}
	\end{array} \right)},
\label{andreevH}
\eeq
where $E^{(+)} = E + \frac{\dmu}{2}$ and $\delta\mu = \mu_{\uparrow} - \mu_{\downarrow}$.
Another set of equations, for components $\utm{r}, \vtp{r}$, one obtains
by replacing $\pair{r}\rightarrow -\pair{r}$ and $E^{(+)}\rightarrow E^{(-)}$, where
$E^{(-)} = E - \frac{\dmu}{2}$.
Thus, within the pure Andreev approximation (ie. neglecting small deviations of
quasiparticle energies from the Fermi energy), each particle is exactly retroreflected
as a hole, and the above equation describes, in practice, the family of one-dimensional problems
associated with each trajectory. 
%&&&&&&&&&&&&&&&&&&&&&&&&&&&&&&&&&&&&&&&&&&&&&&&&&&
\begin{figure}[t]
   \begin{center}
   \includegraphics[width=\columnwidth]{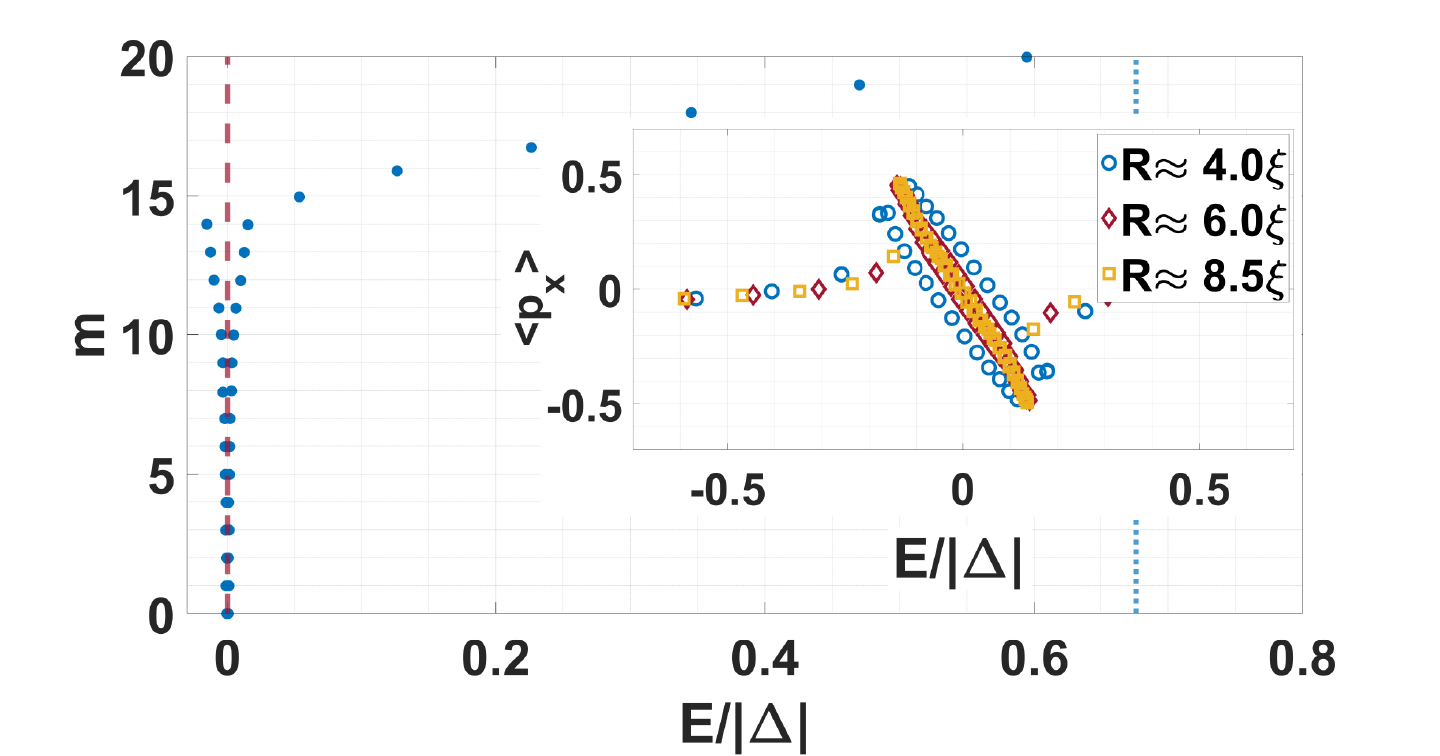} 
   \end{center}\vspace{-3mm}
   \caption{An example of a spectrum of Andreev states in the case of no superflow $q=0$
    for a ferron of radius $R\approx8.5\xi$. The bulk strength of the pairing field is   
    $\Delta/\eF = 0.365$.
    The quasiparticle energies are shifted by $\delta\mu/2$, and thus, the branch is centered around zero.
    The dashed vertical line shows $E = 0$ whereas the dotted vertical line points the value of $\delta\mu/2\Delta$.
    In the inset the Andreev states for various ferron sizes are shown in the presence of  
    superflow of velocity $q/v_{\mathrm{F}}=0.03$. 
    }
   \label{fig:ferron spectrum}
\end{figure}
%&&&&&&&&&&&&&&&&&&&&&&&&&&&&&&&&&&&&&&&&&&&&&&&&&&
For the pairing field of the form (\ref{imprint})
eq. (\ref{andreevH}) provides a quantization condition along each trajectory
of length $L$ which reads
\begin{equation}
\frac{E^{(\pm)}_{n} L}{k_{\mathrm{F}}} - \arccos{\left ( \frac{E^{(\pm)}_{n}}{\Delta} \right ) } + \frac{1}{2}\delta\varphi = \pi n,
\label{Andreevlevels}
\end{equation}
where $n$ is an integer and $\delta\varphi$ denotes the difference of phases of the pairing field $\pair{r}$ between the points at which retro reflections occur. $L$ is the length of the trajectory which the particle or hole follows during the retro reflections.
Since $L$ may vary continuously, the above formula provides a continuous set of 
solutions parameterized by the length of the trajectory.
In the case of no superflow ($q=0$) the trajectory between 
inner $(R_{\mathrm{in}})$ and outer $(R_{\mathrm{out}})$ areas correspond to particle or hole bouncing
back and forth inside the pairing potential of phase difference $\delta\varphi=\pi$.
One solution which always exists in such a case correspond to $E^{(\pm)} = 0$ ($n=0$) and
gives rise to two degenerate branches (if we neglect tunneling
through the ferron interior) with energies $E_{\pm} = \pm\frac{1}{2}\delta\mu$. 
Since the quantization condition given by Eq.(\ref{Andreevlevels}) assumes retro reflection
and does not provide a quantization condition for the transverse particle or hole motion the number
of states in these branches cannot be deduced from this equation. 
The number of states can be found by associating with each trajectory
quantum numbers related to angular momentum. This quantity is conserved
in retro reflection and the quantization of angular momentum provides information
about the number states.
Note that Eq. (\ref{Andreevlevels}) provides the
same result irrespective of dimensionality of the problem.
Indeed the only difference between 2D (circular) and 3D (spherical) ferrons is related
to the density of states, due to the fact that in the 2D case the number of states
in each branch corresponds to the number of angular momentum eigenstates $L_{z}$, 
spanned between $\pm \kF R_{\mathrm{in}}$, whereas in the 3D case, the number of states correspond to
orbital quantum number varying between $0$ and $\kF R_{\mathrm{in}}$ with additional
$2l+1$ degeneracy.
A typical spectrum of the 2D ferron, obtained by solving a full BdG equation, can be seen in Fig. \ref{fig:ferron spectrum}.
The single branch of nearly degenerate states corresponding to
$E_{+} \approx \frac{1}{2}\delta\mu$ is clearly visible.
A small fraction of sub-gap states with absolute values
of angular momenta exceeding $\kF R_{\mathrm{in}}$ but
less than $\kF R_{\mathrm{out}}$ is seen as having energies 
strongly dependent on angular momentum.

In the case of imposed superflow, the situation is more complicated since
the phase difference depends now on the orientation of a particular trajectory
with respect to the direction of the superflow.
Nevertheless, the phase differences vary between $\delta\varphi_{-}=\pi - 2 q L$
and $\delta\varphi_{+}=\pi + 2 q L$.
Solving eq. (\ref{Andreevlevels}) for these two limiting values $\delta\varphi_{\pm}$
under assumption that $|E/\Delta| \ll 1$ one gets for positive energy solutions,
\begin{equation} \label{splitting}
E_{+} \approx \frac{1}{2}\delta\mu \pm \alpha(L) k_{\mathrm{F}}q,
\end{equation}
where $\alpha(L)=\frac{1}{\frac{1}{2} + \frac{\xi}{L}}$ 
($\xi=\frac{\kF}{2\Delta}$). Therefore, one expects that 
the influence of the superflow on Andreev states will lead to splitting of initial
spectrum $E_{\pm} = \pm\frac{1}{2}\delta\mu$ which grows linearly with superflow velocity $q$.
The numerical simulations presented in Fig.\ref{fig:andreev_states} show that
this estimate works surprisingly well even for relatively large values of $q$
when the ferron is on the verge of instability. It can be also seen, in the inset
of Fig.\ref{fig:ferron spectrum}, that the amount of splitting does not
depend on the ferron size. Namely, in the inset one can see Andreev states
for various sizes of ferrons in the presence of superflow which was set
to $q/v_{\mathrm{F}}=0.03$. Clearly the slope represented by states around $E = 0$
is practically the same for various sizes of impurity. Therefore, it is concluded
that the coefficient $\alpha$ in eq. (\ref{splitting}) is weakly dependent on the ferron size.

\section{Details of 3D time-dependent simulations} \label{appE}
We start from the initial solution for unpolarized, unitary Fermi gas. Subsequently, to create local polarization we apply the spin-selective, time-dependent external Gaussian potential given in \eqn{vt}. Calculations are executed on the spatial lattice of size $68\kFi\times40\kFi\times40\kFi$ in the $xyz$ directions with periodic boundary conditions, and $\kF=(3\pi^2(n_\uparrow + n_\downarrow))^{1/3}\approx1$. $A(t)$ is the time-dependent amplitude of the potential and has the following form:

\begin{equation}
A(t)=\left\lbrace 
\begin{array}{ll}
 A_0\,s(t,t_{\textrm{on}}), & 0\leqslant t<t_{\textrm{on}},\\
 A_0, & t_{\textrm{on}} \leqslant t<t_{\textrm{hold}},\\
  A_0\,[1-s(t-t_{\textrm{hold}},t_{\textrm{off}}-t_{\textrm{hold}})], & t_{\textrm{hold}}\leqslant t<t_{\textrm{off}},\\
  0, & t\geqslant t_{\textrm{off}},
\end{array}
\right. 
\label{at}
\end{equation}
where  $s(t, w)$ denotes the function which smoothly varies from 0 to 1 within time interval $[0, w]$,
\begin{equation}
 s(t,w)=\dfrac{1}{2}+
 \dfrac{1}{2}\tanh\left[\tan\left(  \frac{\pi t}{w}-\frac{\pi}{2} \right) \right].
 \label{eq:switch}
\end{equation}
$A_0$ denotes the amplitude of the potential, which we set to be about $A_0\approx 2\eF$.

\begin{figure}[t]
   \begin{center}
   \includegraphics[width=\columnwidth]{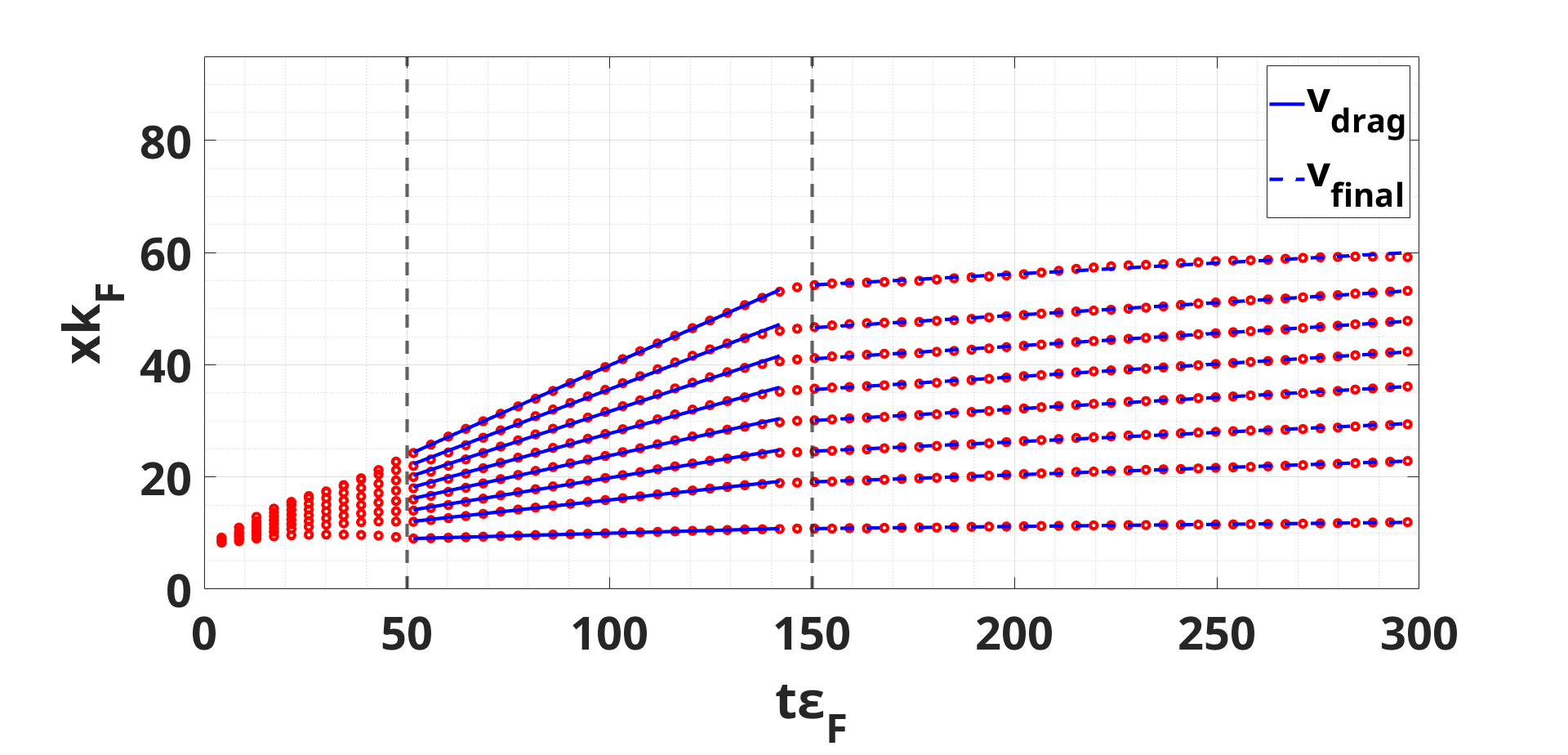} 
   \end{center}\vspace{-3mm}
   \caption{The position of the moving ferron inside a box corresponding to lattice size $68\times40\times40$ which corresponds to $53\xi\times31\xi\times31\xi$ where $\xi$ is the coherence length. The width of the polarizing potential is $\sigma\kF=6$ and its amplitude is $A_0=2\eF$. 
The potential is switched on at $t\eF=50$ and completely removed at $t\eF=150$. Different data sets correspond to different dragging velocities.}
   \label{fig:dragging_ferron}
\end{figure}

To drag the ferron, we set the potential in motion by using $\vdrag \neq 0$ in \eqn{vt}, where $x_0$ is the initial position of the center of the Gaussian potential along the $x$ axis. We extract the velocity 
with which the ferron travels on its own ($\vfin$) by following the position of the center of the polarized sphere. In \fig{fig:dragging_ferron} we provide an example for a potential width $\sigma\kF = 6$. During the switching on the potential, the polarized sphere experiences an acceleration and the potential creates a force responsible for breaking the Cooper pairs. After the potential reaches its maximum amplitude, it is kept on until the nodal sphere is formed. We then turn the potential off and observe the moving impurity.

As $\vdrag$ increases, $\vfin$ eventually reaches a critical value beyond which the ferron can not be accelerated further (Fig.~5). If $\vdrag$ is increased, even more, we observe that the ferron is destroyed during its movement. There are two effects responsible for this: When the final velocity gets closer to the critical value, the ferron undergoes deformation and finally ceases to exist. Moreover, during the acceleration of the ferron, the external potential excites phonons in the system. These phonons scatter inside the simulation box and interact with the ferron. Although for low dragging velocities the ferron is stable against these perturbations, for high velocities the strength of the perturbation increases with the number of excited phonons and eventually the ferron loses its stability. This effect hastens the destruction of the ferron.   

The numerical simulations with the presence of a vortex are conducted at the unitary 
limit. For these calculations we have used a box with the lattice size of $80\times80\times32$ which corresponds to $62\xi\times62\xi\times25\xi$ with $\kF\approx1$. A straight vortex line along the $z$ direction is obtained 
by imposing on the static solution the following structure of the pairing field: $\Delta(x,y) = |\Delta(x,y)| e^{(i\tan^{-1}(y/x))}$. Next, the ferron is generated dynamically by applying the spin selective potential~(\ref{vt}) with $v=0$ and $x_0$ controls the distance of the ferron from the vortex core.

In addition to the results presented in the main article, we present in the Supplemental Material~\cite{supplemental} the dynamics of the ferron placed at the center of the vortex. The movie shows that the polarization that forms the ferron is absorbed into the vortex.

\end{document}

%% file: newcommands.tex
\newcommand{\beq}{\begin{equation}}
\newcommand{\eeq}{\end{equation}}
\newcommand{\bea}{\begin{eqnarray}}
\newcommand{\eea}{\end{eqnarray}}
\newcommand{\eqn}[1]{Eq.~(\ref{#1})}
\newcommand{\fig}[1]{Fig.~\ref{#1}}
\newcommand{\eF}{\varepsilon_{\textrm{F}}}

\newcommand{\kF}{k_{\textrm{F}}}
\newcommand{\kFi}{k_{\textrm{F}}^{-1}}

\newcommand{\grad}{\boldsymbol{\nabla}}

\newcommand{\tbtham}[4]{\left( \begin{array}{cc}
#1 & #2 \\ #3 & #4
\end{array} \right)}

\newcommand{\dmu}{\delta\mu}

\newcommand{\pair}[1]{\Delta(\boldsymbol{#1})}

\newcommand{\abspair}[1]{|\Delta_{#1}|}

\newcommand{\utp}[1]{\tilde{u}_{\uparrow}(\boldsymbol{#1})}
\newcommand{\vtp}[1]{\tilde{v}_{\uparrow}(\boldsymbol{#1})}
\newcommand{\utm}[1]{\tilde{u}_{\downarrow}(\boldsymbol{#1})}
\newcommand{\vtm}[1]{\tilde{v}_{\downarrow}(\boldsymbol{#1})}

\newcommand{\vcrit}{v_{\textrm{crit}}}
\newcommand{\vdrag}{v_{\textrm{drag}}}

\newcommand{\vfin}{v_{\textrm{final}}}